\documentclass[12pt,english]{article}
\usepackage[utf8]{inputenc}
\usepackage[english]{babel}
\usepackage{float,caption,subcaption}
\usepackage{booktabs,longtable}
\usepackage{amsmath}
\usepackage{amssymb,amsthm,amsfonts}
\usepackage{enumerate}
\usepackage{tikz}
\usetikzlibrary{arrows, arrows.meta, positioning, calc}
\usepackage[authoryear,round]{natbib}
\usepackage[hyphens]{url}
\usepackage{hyperref}
\usepackage[a4paper,body={16cm,23cm}]{geometry}

\hypersetup{colorlinks=true, linkcolor=cyan, citecolor=cyan, urlcolor=black, breaklinks=true}
\newtheorem{theorem}{Theorem}
\newtheorem{lemma}{Lemma}

\newtheorem{remark}{Remark}
\newtheorem{example}{Example}

\begin{document}

\title{\textbf{On the fair abatement of riparian pollution}\thanks{We thank 
Erik Ansink and Jens Gudmundsson for helpful comments and suggestions. Financial support from grants PID2023-147391NB-I00 and PID2023-146364NB-I00, funded by MCIU/AEI/10.13039/501100011033 and FSE+, is gratefully acknowledged.
}}
\author{\textbf{Ricardo Mart\'{\i}nez}\thanks{Universidad de Granada. Departamento de Teor\'{\i}a e Historia Econ\'{o}mica. Campus Universitario de La Cartuja s/n. 18011.  Granada, Spain. e-mail: ricardomartinez@ugr.es}$\qquad$ \textbf{Juan
D. Moreno-Ternero}\thanks{Corresponding author. Universidad Pablo
de Olavide. Department of Economics. Carretera de Utrera, Km. 1. 41013. Seville, Spain. e-mail: jdmoreno@upo.es} \quad } \maketitle

\begin{abstract}
  We study the design of fair allocation rules for the abatement of riparian pollution. To do so, we consider the so-called river pollution claims model, recently introduced by \cite{yang2025pollute} to distribute a budget of emissions permits among agents (cities, provinces, or countries) located along a river. In such a model, each agent has a claim reflecting population, emission history, and business-as-usual emissions, and the issue is to allocate among them a budget that is lower (or equal) than the aggregate claim. For environmental reasons, the specific location along the river where pollutants are emitted is an important concern (the more upstream the location is the higher the damage of polluting the river). We characterize a class of \textit{geometric} rules that adjust proportional allocations to compromise between fairness and environmental concerns. Our class is an alternative to the one proposed by \cite{yang2025pollute}. We compare both alternatives through an axiomatic study, as well as an illustration for the case study of the Tuojiang Basin in China.
\end{abstract}

\medskip{}

\medskip{}  \medskip{}

\newpage


\section{Introduction}
Water pollution is becoming a serious environmental problem worldwide, for which effective control measures are urgent \citep[e.g.,][]{Chakraborti2022, Xie2023}. 
The US Environmental Protection Agency launched in 2024 regulations for effluent limitations guidelines and standards for the steam electric power generating category, aimed at reducing pollutants discharged through wastewater from coal-fired power plants by approximately 660 million to 672 million pounds per year.\footnote{\url{https://www.epa.gov/eg/steam-electric-power-generating-effluent-guidelines-2024-final-rule}. This was nevertheless partially reversed a year later: \url{https://www.epa.gov/eg/steam-electric-power-generating-effluent-guidelines-deadline-extensions-rule}.} 
Around the same time, the EU strengthened treatment rules for more thorough and cost-effective urban wastewater management.\footnote{\url{https://environment.ec.europa.eu/news/new-rules-urban-wastewater-management-set-enter-force-2024-12-20_en}} 
In China, where water pollution has become a more serious environmental problem than water scarcity \citep[e.g.,][]{Liu2005}, the Water Pollution Prevention and Control Law was last amended in 2017.\footnote{\url{https://english.mee.gov.cn/Resources/laws/environmental_laws/202012/t20201211_812662.shtml}. 
As our theoretical analysis will partially be motivated by a case study for the Tuojiang Basin in China, it seems relevant to provide some context about current policies in China. That is why we bring some excerpts of this law to Appendix B.}

In this paper, we are concerned with the abatement of riparian pollution. 
As many rivers flow through different provinces, regions, and even countries, a goal is to formalize allocation rules (for the distribution of water pollution among them) that are based on solid normative grounds and determine the amount each agent is allowed to pollute. 
\cite{Ambec02} is the seminal contribution to analyze the allocation of water that flows along a river among the agents 
located therein.\footnote{See also \cite{Brink2012, beal2013, Gudmundsson2019, Oeztuerk2020} and \cite{Wang2022}, among others. \cite{Ansink12, Ansink15} extend this stylized model to account for agents' claims on the river water too. See also \cite{EstevezFernandez21}.} \cite{Ni2007} studied the somewhat dual problem of cleaning a polluted river, in which the input is the cost associated with each agent located along the river.\footnote{See also \cite{Dong2012} and \cite{AlcaldeUnzu2015, AlcaldeUnzu2020} among others.} Now, more recently, 
\cite{yang2025pollute} have introduced another stylized model to analyze the problem of distributing a budget of pollution permits among agents located along a river, which will be our focus in this paper. 
This \textit{river pollution claims problem} extends the so-called claims problem to the case where claimants are located along a line, from upstream to downstream, reflecting the direction of river flow.\footnote{The claims problem, first formalized by \cite{ONeill1982}, refers to one of the oldest problems in the history of economic thought. \cite{Thomson2019} is an excellent survey of the sizable related literature.} Note that the specific location along the river where pollutants are emitted is an important concern, as the higher the location upstream, the greater the damage caused by polluting the river. 

\cite{yang2025pollute} characterize a family of allocation rules for river pollution claims problems (dubbed \textit{externality-adjusted proportional rules}) that convey a balance between fairness and environmental concerns. More precisely, the family ranges from the \textit{proportional} rule, which assigns each agent a proportional amount of the budget to the agent's claim, thus ignoring agents' locations along the river, to the \textit{full transfer} rule, which assigns the whole budget to the most downstream agent, thus ignoring agents' claims. The family of rules is actually defined via averaging between these two extreme rules. The higher the weight for the proportional rule (in the averaging process), the greater the concern for fairness of the resulting rule. The higher the weight for the full transfer rule (in the averaging process), the higher the environmental concern of the resulting rule. 

In this paper, we provide a different family of allocation rules that also offer compromises between fairness and environmental concerns. In our case, this compromise is
obtained by imposing concatenated transfers of residual claims downstream. More precisely, we assume that the most upstream agent retains a portion ($\gamma\in[0,1]$) of its proportional right to the bundle. The residual amount (from this agent's claim) is transferred down to the immediate successor along the river (the second most upstream agent). This agent then increases its claim, adding this transfer to its original claim, and retains the same portion ($\gamma\in[0,1]$) of its (subsequently augmented) proportional right to the bundle. The process continues downstream, \textit{bubbling down} claims until the most downstream agent gets the residual from the budget.\footnote{This process is inspired by the \textit{bubbling up} transfer rules for revenue sharing in hierarchies introduced by \cite{Hougaard17}. See also \cite{hougaard2022optimal} and \cite{Gudmundsson2025}.} When the portion is null ($\gamma=0$), we obtain the full transfer rule. When the portion is full ($\gamma=1$), we obtain the proportional rule. When the portion is between those values ($\gamma\in(0,1)$) we obtain rules that compromise among the extreme rules in a systematically different way than the family of externality-adjusted proportional rules introduced by \cite{yang2025pollute}. The parameter describing the family ($\gamma$) does not represent some abstract compromise, but the actual transfer of some share of \textit{pollution quota} to the next downstream agent. For each country/region, the parameter $\gamma\in[0,1]$ would signal how much it would gain from upstream and how much it would have to give up to benefit downstream. 


We characterize the family of \textit{geometric} rules described above combining several axioms,  
which are all formally introduced (as well as properly discussed) later in the text. We also characterize the (more general) family of rules arising from the resulting set of axioms after dismissing one of them. 
Going in the opposite direction, we characterize the two extreme (and somewhat polar) members of the family (the proportional and full-transfer rules) adding two extra axioms, 
respectively.
We also provide an illustration of how our rules work for a case study of the Tuojiang Basin in China. This, together with the axiomatic study, allows us to properly compare our family with the family of externality-adjusted proportional rules. 

We conclude this introduction discussing some caveats of our model to study the river pollution claims problem, as well as defending the practical implications of our analysis.  


First, our model is formulated theoretically as involving \textit{agents} located along the river. This somewhat flexible term allows to accommodate several interpretations. One might be \textit{permitted polluters}, which is perhaps more accurately connected to actual water pollution control policy implementation. Another could be \textit{cities}, as in our case study. 
And yet another could be \textit{countries}, which would render our analysis a theoretical exercise to deal with the issue of transboundary pollution concerns. Needless to say, the latter is an issue that requires a more general approach, involving (among other aspects) international bargaining protocols, which are absent from our analysis.  

Second, a key assumption of the model is that upstream polluters produce more environmental damage as a result of their location along the river. One might 
argue that this ignores the effect of dilution unless claims are interpreted to be expressed in net terms, thus accounting for dilution too.\footnote{Dilution is usually measured by the ratio of flow in the stream receiving discharge to the flow of wastewater discharge \citep[e.g.,][]{Rice2017}.} 
Now, one might want to enrich the model to also account for the flow at each location, which would permit the computation of the dilution factor at each location.\footnote{This might be similar to account for the length of the river as a factor in the analysis, as one might interpret that a longer distance between one agent and its upstream agent is reflected with a larger flow.} The rules we present in our model could potentially be extended to the enriched framework, although the normative principles on which they would rest might be different.\footnote{This is reminiscent of the extension of claims problems to account for baselines \citep[e.g.,][]{Hougaard2012, Hougaard2013}.}  Alternatively, one could simply perform an ex-post analysis of the rules in our model computing the associated dilution factor to each of them. 

Third, our model is framed as a benchmark case in which rivers have linear structures. We nevertheless illustrate how to extend the rules we obtain in this benchmark case to accommodate different (more general) spatial frameworks, which might account for the existence of tributaries, lakes, reservoirs or other peculiarities of the river basins.

The rest of the paper is organized as follows. In Section \ref{model}, we describe the model 
and all the basic concepts we need for our analysis. 
In Section \ref{main}, we provide our axiomatic study. 
In Section \ref{illustration}, we provide our case study for the Tuojiang Basin in China. In Section \ref{extensions}, we develop possible extensions of our model. Finally, we conclude in Section \ref{discussion}. 
For a smooth passage, we have deferred some technical proofs and additional details of the analysis to Appendix A. Finally, to complement our case study, some relevant excerpts from the Chinese legislation are gathered in Appendix B. 

\section{The model}\label{model}
We consider the same model introduced by \cite{yang2025pollute}. 
Let $\mathbb{N}$ represent the set of all potential \textbf{agents} (cities, regions, countries, or simply permitted polluters). Each river involves a specific set of agents $N=\left\{ 1,...,n\right\}\subset \mathbb{N}$ (sometimes referred in the text as population), which implies a linear structure. By $i < j$ we mean agent $i$ is upstream of agent $j$. 

For each $i \in N$, there is a \textbf{claim} $c_i \geq 0$ capturing the historical pollution of agent $i$ on the river. We interpret a zero claim as a river location where nonpoint source pollution occurs, whereas a positive claim will refer to point source pollution. 
Let $c=\left(c_1, \ldots, c_n\right) \in \mathbb{R}^n_+$ be the \textbf{claims profile} and $C=\sum_{i=1}^n c_i>0$ be the aggregate claim.\footnote{For each $z \in \mathbb{R}^n_+$, we denote by $z_{-i}$ the vector resulting from removing the component $z_i$ from $z$.} That is, we assume the river suffers some point source pollution.  
Let $E\le C$ be the total pollution that society is willing to accept, dubbed the \textbf{budget}. A river pollution abatement problem (in short, a \textbf{problem}) is $(c,E)$. Let $\mathcal{D}^{N}$ denote the domain of problems involving population $N$ and $\mathcal{D}\equiv\bigcup_{N\subset \mathbb{N}}\mathcal{D}^{N}$.
 
\textbf{Redistribution problems} are those in which budget and aggregate claim coincide. That is, society accepts the pollution burden but allows for some redistribution of permits via quotas. Formally, $\mathcal{R}^{N}\equiv\{(c,E)\in \mathcal{D}^{N}: C=E\}$, and $\mathcal{R}\equiv\bigcup_{N\subset \mathbb{N}}\mathcal{R}^{N}\subset \mathcal{D}$. 

Our aim is to provide rules that associate with each problem a balanced \textbf{allocation} of non-negative pollution quotas, i.e.,  $x=\left(x_1, \ldots, x_n\right) \in \mathbb{R}^n_+$ is such that $\sum_{i=1}^n x_i=E$. A \textbf{rule} $\varphi$ is thus a function defined over $\mathcal{D}$ that associates with each $N\subset \mathbb{N}$ and each $(c,E)\in \mathcal{D}^{N}$ an allocation $\varphi(c,E)\in \mathbb{R}^n_+$. We stress that, in contrast to \cite{yang2025pollute}, we do not impose from the outset that rules satisfy \textit{continuity}. 

A basic example is the rule assigning the most downstream agent the whole budget. The rationale is to minimize environmental damage (as no other agent along the river is allowed to pollute). 

\textbf{Full transfer rule}. For each $(c,E) \in \mathcal{D}^{N}$,
$$
\varphi^{FT}(c,E) = \left(0,\dots,0, E\right).
$$

An alternative is the rule that assigns each agent a proportional amount of the budget, thus dismissing concerns over environmental damage (but imposing the old-standing principle of fairness that can be traced back to Aristotle). 

\textbf{Proportional rule}. For each $(c,E) \in \mathcal{D}^{N}$, and each $i\in N$,
$$
\varphi_i^{P}(c,E) = \frac{E}{C}c_i.
$$

A natural compromise is to average those allocations, giving rise to the \textit{externality-adjusted proportional rules}, introduced by \cite{yang2025pollute}.\footnote{They are also reminiscent of the \textit{fixed fraction rules} introduced by \cite{Gudmundsson2024} to share sequentially triggered losses.} We refer to them here as \textit{averaging} rules. 

\textbf{Averaging rules}. For each $\lambda\in[0,1]$, and each $(c,E) \in \mathcal{D}^{N}$,
$$
\varphi^{\lambda}(c,E) =\lambda \varphi^{P}(c,E) +(1-\lambda) \varphi^{FT}(c,E).
$$
Notice that when $\lambda=0$ we obtain the full transfer rule and when $\lambda=1$ we obtain the proportional rule, i.e., $\varphi^0\equiv \varphi^{FT}$, and $\varphi^1\equiv \varphi^{P}$. 


An alternative compromise is obtained with the following family of \textit{geometric rules}, which will be the object of our study. 
In words, geometric rules grant the most upstream agent a portion of its proportional share of the budget. The remaining part of this agent's claim is \textit{bubbled down} to the second most upstream agent, who retains the same portion of its consequently augmented proportional share. The same process continues until the most downstream agent, who simply retains the residual of the budget. Formally,

\textbf{Geometric rules}. For each $\gamma \in[0,1]$ and each $(c,E) \in \mathcal{D}^{N}$, 
$$
\varphi^\gamma_i(c,E) = \gamma  \left(  c_i +  \sum_{k=1}^{i-1} (1-\gamma)^{i-k}c_k \right)\frac{E}{C},
$$
for each $i \in \{1,\dots,n-1\}$, and
$$
\varphi^\gamma_n(c,E) =  \left(c_n +  \sum_{k=1}^{n-1} (1-\gamma)^{n-k} c_k\right)\frac{E}{C}.
$$

Notice that when $\gamma=0$ we obtain the full transfer rule and when $\gamma=1$ we obtain the proportional rule, i.e., $\varphi^0\equiv \varphi^{FT}$, and $\varphi^1\equiv \varphi^{P}$. That is, as in the case of the averaging rules, the family ranges from the full transfer rule to the proportional rule, encompassing a wide variety of compromises, modeled by the parameter $\gamma\in[0,1]$. 

\begin{remark}
With geometric rules, downstream agents may possibly receive more than their claims, as part of their upstream agents’ claims are bubbled down. In other words, geometric rules may violate \textit{claims-boundedness}, which is in stark contrast with the literature on claims problems \citep[e.g.,][] {Thomson2019}. As argued by \cite{yang2025pollute}, this may even be desirable in some cases (for instance, when there is only a small reduction in the number of permits to allocate compared to historic levels, and then the focus might be more on shifting pollution from upstream to downstream).\footnote{The possibility that polluting sources deliberately exceed the regulations in stock pollution problems has recently received attention in the literature \citep[e.g.,][]{Arguedas2020}.} 
\end{remark}

\begin{example} \label{example_geometric}
We illustrate the computation of $\varphi^{0.5}(c,E)$, for $(c,E)=((2,5,5,3), 5)$. As the budget is one third of the aggregate claim, i.e., $\frac{E}{C}=\frac{1}{3}$, each agent $i \in \{1,\ldots,n-1\}$ gets  $\frac{1}{3}\gamma r_i $, where $r_i= \left(  c_i +  \sum_{k=1}^{i-1} (1-\gamma)^{i-k}c_k \right)$ is $i$'s (geometrically) augmented claim.  

Thus, agent 1 retains one third of half its claim: $\frac{1}{3}\gamma c_1=\frac{1}{3}$. 
Agent 2 increases its claim with the remaining half claim from agent 1, i.e., $(1-\gamma )c_1=1$, and retains one third of half this augmented claim: $\frac{1}{3}\gamma(5+1)=1$.
Agent 3 increases its claim with the amount received from upstream, i.e., $(1-\gamma)6=(1-\gamma)5+(1-\gamma)^2 2=3$, and retains one third of half this augmented claim: $\frac{1}{3}\gamma(3+5)=\frac{4}{3}$. 
Finally, agent 4 gets the residual amount: $5-\frac{1}{3}-1-\frac{4}{3}=\frac{7}{3}$. Therefore, the allocation is
$$
\varphi^{0.5}(c,E) = \left( 1 \cdot \frac{E}{C}, 3 \cdot \frac{E}{C}, 4 \cdot \frac{E}{C}, 7 \cdot \frac{E}{C} \right) = \left(\frac{1}{3}, 1, \frac{4}{3}, \frac{7}{3} \right).
$$
\end{example}

To conclude, note that agents with zero claim (interpreted as nonpoint source pollution in our model) play an interesting role for geometric rules. 
More precisely, they hurt point source pollution locations downstream, which would receive a higher quota with point source pollution locations upstream. That is, a positive claim is partially transferred downstream with a geometric rule, whereas, obviously, a null claim cannot transfer any part of it. 

\section{Axioms and characterizations}\label{main}

We introduce some axioms that formalize principles with normative appeal for this setting.

The first axiom we consider states that, if all claims and the budget are multiplied by the same positive real number, the allocation does so too. This requirement can be interpreted in the following way. If we change the units in which inputs are measured (from liters to gallons, for example), the units of the allocation must change accordingly. This is a standard axiom in the literature \citep[e.g.,][]{moulin2000priority, Thomson2019}. Note that, in practical terms, this axiom excludes the use of \textit{threshold} rules, which would allocate pollution quotas differently depending on whether the budget is below or above a certain predetermined threshold. 

\textbf{Scale invariance}. For each $(c,E) \in \mathcal{D}$ and each $\mu \in \mathbb{R}_+$, $\varphi(\mu c,\mu E)=\mu \varphi(c,E)$.

The second axiom was considered by \cite{yang2025pollute} in this same context, and it also has a long tradition of use in axiomatic work for other models of resource allocation \citep[e.g.,][]{chun1988proportional, Thomson2019}. It states that if the budget comes in two installments, the allocation chosen for the sum of these two installments should coincide with the sum of the allocations chosen for each of them. For instance, in practical terms, the allocation could be unveiled by the government once or twice a year (assuming the overall amount to reduce is the same) and that choice would not have an effect on the process.  

\textbf{Budget additivity}. For each $(c, E) \in \mathcal{D}$ and each pair $E', E'' \geq 0$ such that $E=E'+E''$,
$$
\varphi(c, E)=\varphi (c, E')+ \varphi (c,E'').
$$

The remaining axioms we consider apply only to the subdomain of redistribution problems $\mathcal{R}$ (that is, when the budget is equal to the aggregate claim). 

The first axiom in this group states that if a claim increases, this change can only alter the allocation of downstream agents.\footnote{This is inspired by the namesake axiom in \cite{Martinez2025} for the allocation of riparian water rights.} Practically, the axiom implies that pollution follows the river stream and, thus, wastewater discharge at a given location typically does not affect those located upstream. This is a realistic scenario with the exception, for instance, in which migratory fish species are affected downstream and subsequently fail to return upstream. 

\textbf{Upstream invariance}. Let $(c,E),(\hat{c},E') \in \mathcal{R}^{N}$ be such that $c_i<\hat{c}_i$ for some $i \in N$, and $c_{-i} = \hat{c}_{-i}$. Then, for each $k<i$,
$$
\varphi_k(c,E) = \varphi_k(\hat{c},E').
$$

The next requirement states that if two rivers only have one point source pollution (that is, all claims except one are zero), and they pollute the same, then they receive the same pollution quota. In practical terms, this axiom refers to the likely scenario in which there exists a single pollution source along the river, such as a wastewater treatment plant, factory, or sewage outlet, which can cause severe environmental damage. When society is willing to accept its pollution (that is, the budget coincides with the aggregate claim and the only concern is to redistribute via pollution quotas), the axiom states the quota for this single point source pollution will be independent of the structure of the river. Imposing no pollution reduction to it (that is, assigning to it a quota equal to the whole claim, and zero to all non-point sources) is an admissible option (albeit not the only one). 

\textbf{Equal treatment of equal single polluters}. For each $E>0$, and each pair $i,j\in\{1,\dots, n\}$,
$$\varphi_i((\overbrace{0, \ldots, 0}^{i-1}, E, \overbrace{0, \ldots, 0}^{n-i}),E) = \varphi_j((\overbrace{0, \ldots, 0}^{j-1}, E, \overbrace{0, \ldots, 0}^{n-j}),E).$$

Finally, an axiom addressing the situation in which the most upstream agent leaves with its assigned permits (provided the amount is not above the initial claim), while transferring the residual claim to the following agent downstream.\footnote{This is reminiscent of the so-called \textit{lowest-rank consistency} axiom in \cite{Hougaard17} for the problem of revenue sharing in hierarchies. It is obviously connected to the long-standing principle of consistency, which has played a critical role in axiomatic work \citep[e.g.,][]{moulin2000priority, thomson2012axiomatics}, and which was also considered in this context by \cite{yang2025pollute}.} The axiom states that if the remaining agents reevaluate the reduced problem (which belongs to the same domain of redistribution problems), their assignments are unchanged. In practical terms, the axiom formalizes a natural stability requirement for the likely scenario in which the allocation of permits might arise as the outcome of a negotiation process (for instance, in the case of international rivers). In that scenario, one might think of a plausible case in which the most upstream agent is a more powerful agent that played a more critical role in the negotiation process. Once that agent is pleased with the negotiation outcome, and leaves the table with the allocation (and a likely residual claim transferred to the next agent downstream), the remaining agents might renegotiate their allocation. The axiom states nothing should change in that case.  

\textbf{Top consistency}. For each $(c,E) \in \mathcal{R}^N$, such that $\varphi_1(c,E) \leq c_1$,
$$
\varphi_{-1}(c,E) = \varphi \left( (c_2+(c_1-\varphi_1(c,E)),c_3,\ldots,c_n), E-\varphi_1(c,E) \right).
$$

Prior to presenting our characterizations, we introduce a lemma that shows a noteworthy implication of \emph{additivity} within our framework. More specifically, this axiom enables the extension of characterizations from the restricted domain of redistribution problems ($\mathcal{R}$) to the unrestricted universal domain of problems ($\mathcal{D}$). 

\begin{lemma}\label{lemma}
Let $\varphi$ and $\varphi'$ be two rules that satisfy \emph{budget additivity}. If, for each $(c,E)\in \mathcal{R}$, $\varphi(c,E)=\varphi'(c,E)$, then it follows that for each $(c,E)\in \mathcal{D}$, $\varphi(c,E)=\varphi'(c,E)$.
\end{lemma}

The combination of the previous axioms characterizes the family of geometric rules.

\begin{theorem}\label{thm_geom}
A rule satisfies scale invariance, upstream invariance, equal treatment of equal single polluters, top consistency and budget additivity if and only if it is a geometric rule.
\end{theorem}

We now generalize Theorem \ref{thm_geom} by omitting one of its axioms, thereby characterizing a more general family of rules, which obviously encompasses the geometric rules. More precisely, we dismiss scale invariance, as violations of this axiom might not be uncommon in the specific framework we study. For instance, the boundary water treaty of 1909, which governs the use of waters shared by Canada and the United States, stipulates allocations that contravene scale invariance.\footnote{\url{https://www.ijc.org/en/boundary-waters-treaty-1909}. Last accessed June 3, 2025.} 

As the next result states, if we combine all the axioms from Theorem \ref{thm_geom}, except for scale invariance, we characterize the family of \emph{generalized geometric rules}, whose formal definition is the following.

\textbf{Generalized geometric rules}. For each function $\Gamma:\mathbb{R}_+ \longrightarrow \mathbb{R}_+$ such that $\Gamma(t) \le t$, each $(c,E) \in \mathcal{D}$ and each $i \in N$, 
$$
\varphi^\Gamma_i(c,E) = r^\Gamma_i(c) \frac{E}{C},
$$
where
$$
r^\Gamma_i(c) = 
\begin{cases}
    \Gamma(c_1)  & \text{if } i=1, \\[0.1cm]
    \Gamma\left( c_i + \sum_{k=1}^{i-1} \left( c_k-r^\Gamma_k(c) \right) \right)  & \text{if } i \in \{2,\ldots,n-1\}, \\[0.1cm]
    c_n + \sum_{k=1}^{n-1} \left( c_k-r^\Gamma_k(c,E) \right)  & \text{if } i=n. 
\end{cases}
$$

Notice that if we set $\Gamma(t)=\gamma t$ for some $\gamma \in [0,1]$, then $\varphi^\Gamma \equiv \varphi^\gamma$. Consequently, geometric rules are indeed included in this new family. But when the $\Gamma$ function is not linear we go beyond the benchmark geometric rules.

\begin{theorem}\label{thm_supergeom}
A rule satisfies upstream invariance, equal treatment of equal single polluters, top consistency and budget additivity if and only if it is a generalized geometric rule.
\end{theorem}

We now go in the opposite direction, 
adding two independent axioms and showing that each of them drives towards a characterization of one of the two extreme members of the families. 

First, we consider a standard impartiality axiom.\footnote{Impartiality axioms appear often in the theory of justice \citep[e.g.,][] {moreno2006impartiality}.} 

\textbf{Equal treatment of equal claims}. For each pair $(c,E) \in \mathcal{D}^N$, and each $\{i,j\} \subset N$, if $c_i=c_j$ then $\varphi_i(c,E)=\varphi_j(c,E)$.

Note that, in spite of the resemblance, this axiom is logically unrelated to equal treatment of equal single polluters.\footnote{To be more precise, the former axiom is punctual whereas the latter is relational \citep[e.g.,][] {Thomson2023}.} Nevertheless, replacing one by the other at Theorem \ref{thm_supergeom}, we move from characterizing the whole family of generalized geometric rules to characterizing only one of its members (the proportional rule). This shows that the axiom has stronger implications (at least, when combined with the remaining axioms from Theorem \ref{thm_supergeom}). This is not entirely surprising as, in practical terms, the axiom is ignoring the location of agents along the river for the allocation of pollution quotas. That is, the quota is only determined by the claim each agent has (and the budget society wants to impose to reduce pollution). 

\begin{theorem}\label{thm_PROP}
A rule satisfies upstream invariance, equal treatment of equal claims, top consistency and budget additivity if and only if it is the proportional rule.
\end{theorem}

To conclude, we add the axiom of \textit{additivity}, whose long tradition in axiomatic work can be traced back to \cite{Shapley1953}. 

\textbf{Additivity}. For each pair $(c,E),(\hat{c},E') \in \mathcal{D}^N$,
$$
\varphi(c+\hat{c},E+E') = \varphi(c,E) + \varphi(\hat{c},E') 
$$

This axiom obviously implies budget additivity. 
Thus, its practical implications are stronger. Its logical implications too. As a matter of fact, when replacing the latter at Theorem \ref{thm_supergeom}, it allows to move from characterizing the whole family of generalized geometric rules to characterizing only one of its members (in this case, the full-transfer rule). 

\begin{theorem}\label{thm_NT}
A rule satisfies upstream invariance, equal treatment of equal single polluters, top consistency and additivity if and only if it is the full-transfer rule.
\end{theorem}

\section{Case study}\label{illustration}
In this section, we replicate the case study in \cite{yang2025pollute}, with data from the Tuojiang River Basin (China), to illustrate how the geometric rules perform in this case, while comparing with the performance of the averaging rules in the same example. 

Management and monitoring of wastewater in China is governed by the Water Pollution Prevention and Control Law of 2017, which, as mentioned in the introduction, applies to the prevention and control of pollution of rivers, lakes, canals, irrigation channels, reservoirs and other surface waters and ground waters within the territory. 
The concern has kept growing in recent years and on May 21, 2025, seven departments, including the Ministry of Ecology and Environment, issued the ``Action Plan for the Protection and Construction of Beautiful Rivers and Lakes (2025-2027)".\footnote{\url{https://cwrrr.org/regulation/beautiful-rivers-and-lakes-protection-and-construction-action-plan-2025-2027/}.} 
The Tuojiang River is one of the rivers with the greatest economic impact in the Sichuan Province, in southeastern China. It spans approximately 638 kilometers and flows through the cities of Deyang, Chengdu, Ziyang, Neijiang, Zigong, and Luzhou. As a major tributary of the Yangtze river, it is strictly managed under the Yangtze River Protection Law and the Water Pollution Prevention and Control Action Plan. Key regulations include mandatory total discharge control for pollutants, stringent industrial effluent standards, and penalties for exceeding allowed limits. 
The law also states that relevant provincial-level people's governments shall formulate stricter requirements for total phosphorus emissions control to effectively control total phosphorus emissions.\footnote{\url{https://english.mee.gov.cn/Resources/laws/environmental_laws/202104/t20210407_827604.shtml}.}

As mentioned by \cite{yang2025pollute}, in 2020, the total volume of water discharged into the river had to be limited to 64.3 million cubic meters (MCM), while the historical aggregate discharge of the six cities was 81.24 MCM. Thus, in order to apply our model, it is natural to consider the objective volume as the budget, and the historical discharges of each city as their claims in our model (second column of Table \ref{table1}).

The allocations resulting from the application of some geometric rules (columns 3 to 7) and the corresponding averaging rules (columns 8 to 12) are collected in Table \ref{table1}. To ease comparison, we have considered the same set of parameters for both families: $\lambda, \gamma \in \{0,\frac{1}{4},\frac{1}{2},\frac{3}{4},1\}$. 
By definition, $\varphi^{\gamma=0} = \varphi^{\lambda=0} = \varphi^{FT}$ and $\varphi^{\gamma=1} = \varphi^{\lambda=1} = \varphi^{P}$. Thus, the interesting comparisons come from the intermediate cases. As we can observe, Luzhou, the most downstream city, is allowed to discharge significantly higher volumes of untreated sewage with the averaging rules than with the geometric ones. In contrast, the four intermediate cities are entitled to larger allocations with the geometric rules, particularly when $\gamma = \lambda = 0.5$, the case in which the total available volume is more evenly distributed.

\begin{table}[h!]
  \setlength{\tabcolsep}{4pt}
  \begin{center}
  \begin{tabular}{lrrrrrrrrrrr}
  \toprule
  & & \multicolumn{5}{c}{Geometric rules $\varphi^\gamma$} & \multicolumn{5}{c}{Averaging rules $\varphi^\lambda$} \\
  \cmidrule(r){3-7} \cmidrule(l){8-12}
  City & $c_i$ & $\gamma=0$ & $\gamma=\frac{1}{4}$ & $\gamma=\frac{1}{2}$ & $\gamma=\frac{3}{4}$ & $\gamma=1$ & $\lambda=0$ & $\lambda=\frac{1}{4}$ & $\lambda=\frac{1}{2}$ & $\lambda=\frac{3}{4}$ & $\lambda=1$ \\ 
  \midrule
  Deyang   & 4.17  & 0    & 0.83  & 1.65  & 2.48  & 3.30  & 0    & 0.83  & 1.65  & 2.48  & 3.30  \\
  Chengdu  & 53.98 & 0    & 11.30 & 22.19 & 32.66 & 42.72 & 0    & 10.68 & 21.36 & 32.04 & 42.72 \\
  Ziyang   & 2.13  & 0    & 8.90  & 11.94 & 9.43  & 1.69  & 0    & 0.42  & 0.84  & 1.26  & 1.69  \\
  Neijiang & 3.30  & 0    & 7.33  & 7.27  & 4.32  & 2.61  & 0    & 0.65  & 1.31  & 1.96  & 2.61  \\
  Zigong   & 2.48  & 0    & 5.98  & 4.62  & 2.55  & 1.96  & 0    & 0.49  & 0.98  & 1.47  & 1.96  \\
  Luzhou   & 15.18 & 64.3 & 29.97 & 16.63 & 12.87 & 12.01 & 64.3 & 51.23 & 38.16 & 25.09 & 12.01 \\
  \cmidrule(r){2-12}
 Total & 81.24 & 64.3 & 64.3 & 64.3 & 64.3 & 64.3 & 64.3 & 64.3 & 64.3 & 64.3 & 64.3 \\
  \bottomrule
  \end{tabular}
  \end{center}
  \caption{Application of several (geometric and averaging) rules to the Tuojiang river. \label{table1}}
  \setlength{\tabcolsep}{6pt}
\end{table}

\begin{figure}[h!]
  \centering
  \includegraphics[scale=0.6]{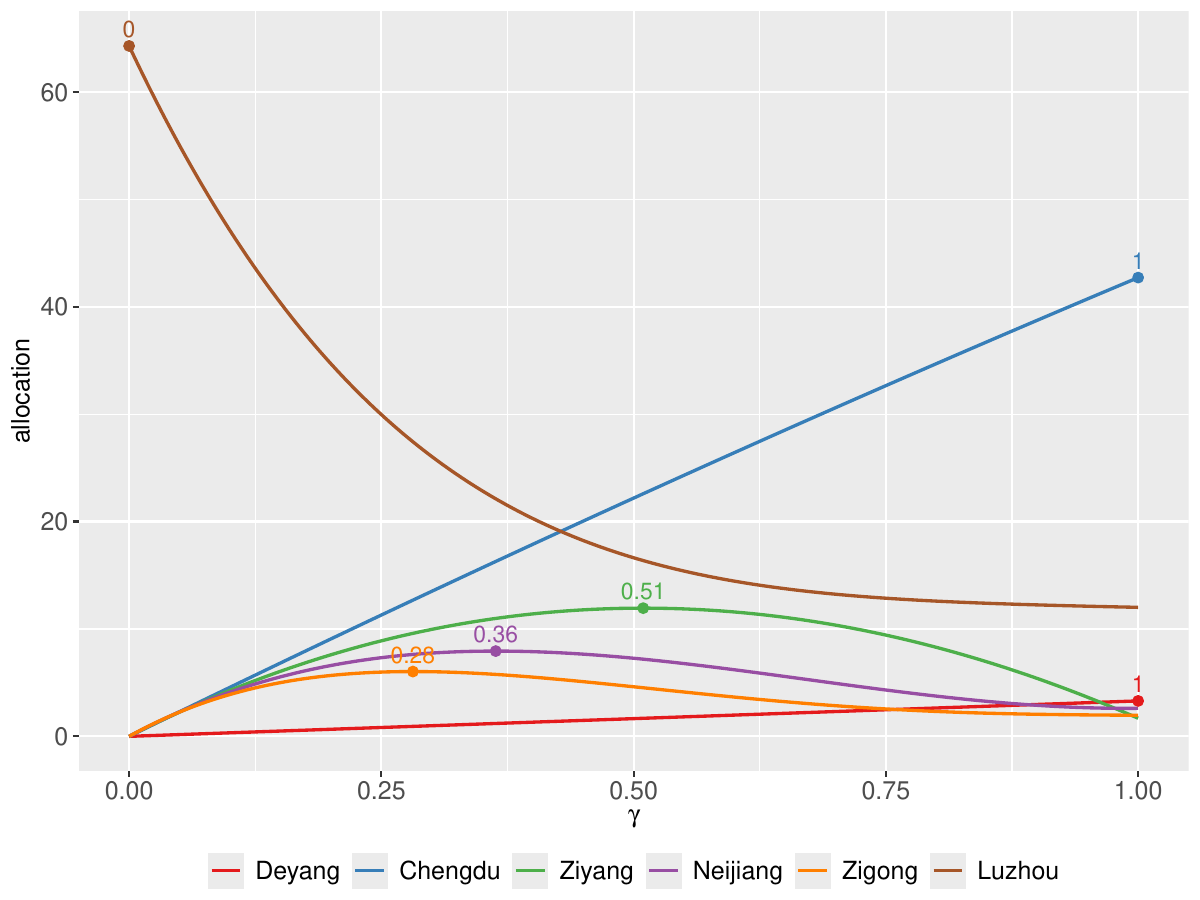}
  \caption{Paths of awards for geometric rules when $\gamma$ varies from $0$ to $1$. \label{fig1}}
\end{figure}


In Figure \ref{fig1}, we plot the granted amounts to each city with the geometric rules, when $\gamma$ varies from $0$ to $1$. We can observe from there that the amount Luzhou (the most downstream city) obtains decreases as $\gamma$ increases. The opposite happens for Deyang and Chengdu (the two most upstream cities). The remaining (intermediate) cities (Ziyang, Neijiang and Zigong) exhibit a different behavior, as they receive higher shares of the budget for intermediate values of $\gamma$. The number above each curve indicates the value of $\gamma$ whose corresponding rule yields the highest amount for that city (within the rules in the geometric family).
\newpage

We also use this case study to address some natural questions emerging from our axiomatic analysis. Why would a planner choose a certain value of $\gamma$ instead of another? Likewise, what are the goals that would drive the decision to set $\gamma$ equal to one value or another? Both Table \ref{table1} and Figure \ref{fig1} provide information that might help answering these questions and guiding towards the selection of $\gamma$ in practice. For example, Table \ref{table1} indicates that some downstream agents receive permits exceeding their claims (a feature we already highlighted for geometric rules at Remark 1). A natural criterion to select the value of $\gamma$ (or, at least, to restrict the feasible range of the values) would be to enforce that no city receives more than its claim (that is, the rule satisfies claims-boundedness). 
It turns out that the minimum value of $\gamma$ that guarantees claims-boundedness in this  case is $\gamma = 0.989$. In other words, based on this case study, one would have a narrow set of options to guarantee claims-boundedness, hence considerably simplifying the task of selecting a rule within the family (or, in other words, the value of $\gamma$). 
Note that an increase in the number of agents does not always reduce the range of values of $\gamma$ that guarantee claims-boundedness for the geometric rules; this may depend on the claims of the incoming agents. To illustrate this aspect, consider for instance the problems $(c,4)$, $(\hat{c},4)$, and $(\tilde{c},4)$, where
$c = (2,2,2)$, $\hat{c} = (2,2,2,1)$, and $\tilde{c} = (2,2,2,3)$. In both $\hat{c}$ and $\tilde{c}$ we add an extra agent at the mouth of the river, but with different claims. The minimum value of $\gamma$ that guarantees claims-boundedness for $\varphi^\gamma(c,E)$ is $\gamma = 0.634$, whereas for $\varphi^\gamma(\hat{c},E)$ and $\varphi^\gamma(\tilde{c},E)$ they are $\gamma = 0.722$ and $\gamma = 0.217$, respectively. Thus, depending on the new agent's claim, the range of $\gamma$ for which the property is satisfied may shrink or expand.

In order to elaborate further on the insights from our case study, we consider an extra variety of hypothetical claim distributions to better establish the nature of the compromises both families offer. To ease the comparison, we keep the set of cities fixed, as well as the budget ($E=64.3$) and the aggregate claim ($C=81.24$), while varying only the distribution of claims across cities. More specifically, we consider the alternative claim vectors $\hat{c}=(4.17,19.89,14,14,14,15.18)$ and $\tilde{c} = (53.98,4.17,2.13,3.30,2.48,15.18)$. 

The choice of $\hat{c}$ was motivated by the fact that, originally, Chengdu has a much larger claim than those of the intermediate cities (Ziyang, Neijiang, and Zigong). Thus, we chose $\hat{c}$ so that those cities would have the same claim, while preserving the aggregate (of the whole claim vector). The allocations for the problem $(\hat{c},E)$ resulting from the application of several geometric rules and the corresponding averaging rules are collected in Table \ref{table4}.\footnote{
Recall that $\varphi^{\gamma=0} \equiv \varphi^{FT} \equiv \varphi^{\lambda=0}$ and $\varphi^{\gamma=1} \equiv \varphi^{FT} \equiv \varphi^{\lambda=1}$. Thus, we leave aside those cases for the comparisons.} As shown in Table \ref{table4}, the geometric and averaging rules display different patterns when claims are identical. That is, under the averaging rules, Ziyang, Neijiang, and Zigong receive the same allocation, regardless of their relative locations along the river. In contrast, location along the river plays a more significant role for the geometric rules, as these three cities receive different amounts depending on that. Moreover, within the geometric family, a higher value of the parameter $\gamma$ reduces these disparities. This additional layer of sensitivity to spatial location is absent for the averaging family.

\begin{table}[h!]
  \setlength{\tabcolsep}{4pt}
  \begin{center}
  \begin{tabular}{lrrrrrrrrrrr}
  \toprule
  & & \multicolumn{5}{c}{Geometric rules $\varphi^\gamma$} & \multicolumn{5}{c}{Averaging rules $\varphi^\lambda$} \\
  \cmidrule(r){3-7} \cmidrule(l){8-12}
  City & $\hat{c}_i$ & $\gamma=0$ & $\gamma=\frac{1}{4}$ & $\gamma=\frac{1}{2}$ & $\gamma=\frac{3}{4}$ & $\gamma=1$ & $\lambda=0$ & $\lambda=\frac{1}{4}$ & $\lambda=\frac{1}{2}$ & $\lambda=\frac{3}{4}$ & $\lambda=1$ \\ 
  \midrule
Deyang & 4.17 & 0.00 & 0.83 & 1.65 & 2.48 & 3.30 & 0.00 & 0.83 & 1.65 & 2.48 & 3.30 \\ 
  Chengdu & 19.89 & 0.00 & 4.55 & 8.70 & 12.43 & 15.74 & 0.00 & 3.94 & 7.87 & 11.81 & 15.74 \\ 
  Ziyang & 14.00 & 0.00 & 6.19 & 9.89 & 11.42 & 11.08 & 0.00 & 2.77 & 5.54 & 8.31 & 11.08 \\ 
  Neijiang & 14.00 & 0.00 & 7.41 & 10.48 & 11.16 & 11.08 & 0.00 & 2.77 & 5.54 & 8.31 & 11.08 \\ 
  Zigong & 14.00 & 0.00 & 8.33 & 10.78 & 11.10 & 11.08 & 0.00 & 2.77 & 5.54 & 8.31 & 11.08 \\ 
  Luzhou & 15.18 & 64.30 & 37.00 & 22.80 & 15.72 & 12.01 & 64.30 & 51.23 & 38.16 & 25.09 & 12.01 \\ 
  \cmidrule(r){2-12}
 Total & 81.24 & 64.3 & 64.3 & 64.3 & 64.3 & 64.3 & 64.3 & 64.3 & 64.3 & 64.3 & 64.3 \\
  \bottomrule
  \end{tabular}
  \end{center}
  \caption{Application of several (geometric and averaging) rules to the hypothetical case $(\hat{c},E)$. \label{table4}}
  \setlength{\tabcolsep}{6pt}
\end{table}

One might also compare both families in this hypothetical configuration $\hat{c}$ yields, with respect to claims-boundedness (i.e., all agents receive permits that do not exceed their claims). 
The minimum value of $\gamma$ that guarantees claims-boundedness for $\varphi^\gamma(\hat{c},E)$ is $\gamma = 0.778$, whereas the minimum value of $\lambda$ that guarantees claims-boundedness for $\varphi^\lambda(\hat{c},E)$ is $\lambda = 0.94$. 

As for $\tilde{c}$, note that it is also a variation of the original distribution $c$. In this case, we exchange the claims of the two most upstream cities; that is, we set $\tilde{c}_\text{Deyang} = c_\text{Chengdu}$ and $\tilde{c}_\text{Chengdu} = c_\text{Deyang}$, while keeping the claims of the remaining cities fixed. The allocations for the problem $(\tilde{c},E)$ resulting from the application of several geometric rules and the corresponding averaging rules are collected in Table \ref{table5}.
It can be seen from there that substantial disparities may arise. 
Note that, at $\tilde{c}$, the claim of the most upstream city (Deyang) is significantly larger than the others'. Under the averaging rules, Deyang retains the $\lambda$ share of its proportional claim, while the remaining portion is assigned entirely to Luzhou, the most downstream city. Under the geometric rules, Deyang also keeps the same fraction $\gamma$ of its proportional claim; however, the remaining portion is distributed among all downstream cities (not only Luzhou), according to their relative positions along the river with respect to Deyang. This explains most of the disparities. 

\begin{table}[h!]
  \setlength{\tabcolsep}{4pt}
  \begin{center}
  \begin{tabular}{lrrrrrrrrrrr}
  \toprule
  & & \multicolumn{5}{c}{Geometric rules $\varphi^\gamma$} & \multicolumn{5}{c}{Averaging rules $\varphi^\lambda$} \\
  \cmidrule(r){3-7} \cmidrule(l){8-12}
  City & $\tilde{c}_i$ & $\gamma=0$ & $\gamma=\frac{1}{4}$ & $\gamma=\frac{1}{2}$ & $\gamma=\frac{3}{4}$ & $\gamma=1$ & $\lambda=0$ & $\lambda=\frac{1}{4}$ & $\lambda=\frac{1}{2}$ & $\lambda=\frac{3}{4}$ & $\lambda=1$ \\ 
  \midrule
Deyang & 53.98 & 0.00 & 10.68 & 21.36 & 32.04 & 42.72 & 0.00 & 10.68 & 21.36 & 32.04 & 42.72 \\ 
  Chengdu & 4.17 & 0.00 & 8.84 & 12.33 & 10.49 & 3.30 & 0.00 & 0.83 & 1.65 & 2.48 & 3.30 \\ 
  Ziyang & 2.13 & 0.00 & 7.05 & 7.01 & 3.89 & 1.69 & 0.00 & 0.42 & 0.84 & 1.26 & 1.69 \\ 
  Neijiang & 3.30 & 0.00 & 5.94 & 4.81 & 2.93 & 2.61 & 0.00 & 0.65 & 1.31 & 1.96 & 2.61 \\ 
  Zigong & 2.48 & 0.00 & 4.95 & 3.39 & 2.20 & 1.96 & 0.00 & 0.49 & 0.98 & 1.47 & 1.96 \\ 
  Luzhou & 15.18 & 64.30 & 26.85 & 15.40 & 12.75 & 12.01 & 64.30 & 51.23 & 38.16 & 25.09 & 12.01 \\ 
  \cmidrule(r){2-12}
 Total & 81.24 & 64.3 & 64.3 & 64.3 & 64.3 & 64.3 & 64.3 & 64.3 & 64.3 & 64.3 & 64.3 \\
  \bottomrule
  \end{tabular}
  \end{center}
  \caption{Application of several (geometric and averaging) rules to the hypothetical case $(\tilde{c},E)$. \label{table5}}
  \setlength{\tabcolsep}{6pt}
\end{table}

Finally, 
note that the minimum value of $\gamma$ that guarantees claims-boundedness for $\varphi^\gamma(\tilde{c},E)$ is $\gamma = 0.977$, whereas the minimum value of $\lambda$ that guarantees claims-boundedness for $\varphi^\lambda(\tilde{c},E)$ is $\lambda = 0.94$. 

\section{Further insights}\label{extensions}
We explore in this section 
how to extend our analysis to consider more intricate hydrological network structures. 

\begin{figure}[H]
\centering
\begin{subfigure}{0.17\textwidth}
\centering
\begin{tikzpicture}[
  node distance=0.6cm and 0.6cm,
  every node/.style={draw, rectangle, minimum width=0.5cm, minimum height=0.5cm, align=center, rounded corners=5pt, font=\footnotesize},
  ->
]
\node (1) {1};
\node (2) [below=of 1] {2};
\node (3) [below=of 2] {3};
\node (4) [below=of 3] {4};
\draw (1) -- (2);
\draw (2) -- (3);
\draw (3) -- (4);
\end{tikzpicture}
\caption{}
\end{subfigure}
\begin{subfigure}{0.25\textwidth}
\centering
\begin{tikzpicture}[
  node distance=0.6cm and 0.6cm,
  every node/.style={draw, rectangle, minimum width=0.5cm, minimum height=0.5cm, align=center, rounded corners=5pt, font=\footnotesize},
  ->
]
\node (1) {1};
\node (2) [below left=of 1] {2};
\node (3) [below right=of 1] {3};
\node (4) [below left=of 3] {4};
\node (5) [below right=of 3] {5};
\node (6) [below =of 5] {6};
\draw (1) -- (2);
\draw (1) -- (3);
\draw (3) -- (4);
\draw (3) -- (5);
\draw (5) -- (6);
\end{tikzpicture}
\caption{}
\end{subfigure}
\begin{subfigure}{0.5\textwidth}
\centering
\begin{tikzpicture}[
  node distance=0.6cm and 0.6cm,
  every node/.style={draw, rectangle, minimum width=0.5cm, minimum height=0.5cm, align=center, rounded corners=5pt, font=\footnotesize},
  ->
]
\node (1) {1};
\node (2) [below left=of 1] {2};
\node (3) [below right=of 1] {3};
\node (4) [below left=of 3] {4};
\node (5) [below =of 4] {5};
\draw (1) -- (2);
\draw (1) -- (3);
\draw (2) -- (4);
\draw (3) -- (4);
\draw (4) -- (5);
\end{tikzpicture}
\caption{}
\end{subfigure}
\caption{\footnotesize{Examples of basin structures: (a) line; (b) hierarchy; (c) boundary river.} \label{fig_structures}}
\end{figure}
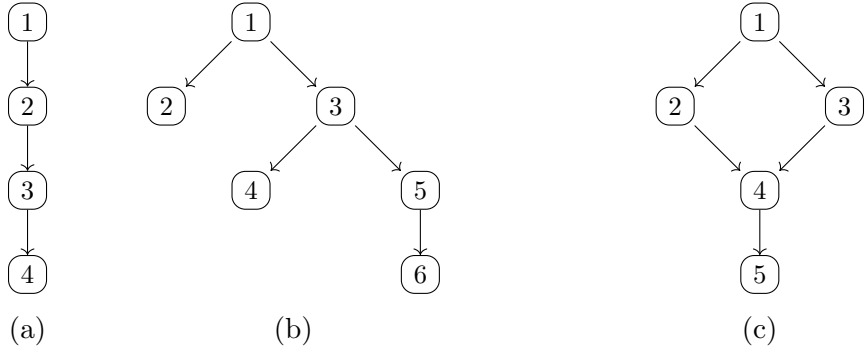

Figure \ref{fig_structures} conveys several structures. Case (a) refers to the benchmark (linear) case we have focused on so far. Case (b) refers to a tree configuration, in which water flows through multiple tributaries that diverge as it progresses downstream. Finally, case (c) presents an even more complex structure, that might refer to a transboundary river. More precisely, in this setting, the river originates in country 1 and diverges in two branches, crossing countries 2 and 3, respectively. Then, the two branches converge again downstream, at country 4, from where it follows downstream until reaching its mouth at country 5. 
\bigskip

Our geometric rules can be adapted to all the configurations shown in Figure \ref{fig_structures}. For ease of comparison, we start providing the expression at case (a), which is precisely the benchmark lineal model. There, for each $i \in \{1,\ldots, 4\}$, $x_i=r_i^a(c) \frac{E}{C}$, where
\begin{align*}
r^a_1(c)&= \gamma c_1,  \\
r^a_2(c)&= \gamma \left[c_2 + (1-\gamma)c_1 \right],  \\
r^a_3(c)&= \gamma \left[c_3 + (1-\gamma)c_2 + (1-\gamma)^2c_1 \right],  \\
r^a_4(c)&= \left[c_4 + (1-\gamma)c_3 + (1-\gamma)^2c_2 + (1-\gamma)^3c_1 \right]. 
\end{align*}
For case (b), and each $i \in \{1,\ldots, 6\}$, the allocation is given by $x_i=r_i^b(c) \frac{E}{C}$, where the proportional shares $r^{b}(c)$ are constructed as follows. Agent 1 retains a fraction $\gamma$ of its claim and transfers the remainder to the two immediate downstream agents, assigning one half to each. Agent 2, as a mouth of the river, simply adds its own claim to the amount received from Agent 1. Agent 3, in contrast, is not a mouth; therefore, to compute its share, its claim is added to the amount transferred from Agent 1, retaining a fraction $\gamma$ of this total. The procedure continues in the same manner downstream until all agents’ proportional shares have been determined. The final shares are presented below.
\begin{align*}
r^b_1(c)&= \gamma c_1, \\
r^b_2(c)&= c_2 + \frac{1}{2}(1-\gamma)c_1, \\
r^b_3(c)&= \gamma \left[c_3 + \frac{1}{2}(1-\gamma)c_1 \right], \\
r^b_4(c)&= c_4 + \frac{1}{2}(1-\gamma)\left( c_3+\frac{1}{2} (1-\gamma)c_1\right) = c_4 + \frac{1}{2}(1-\gamma)c_3 + \frac{1}{4}(1-\gamma)^2c_1,\\
r^b_5(c)&= \gamma \left[c_5 + \frac{1}{2}(1-\gamma)\left( c_3+\frac{1}{2} (1-\gamma)c_1\right) \right]  = \gamma \left[ c_5 + \frac{1}{2}(1-\gamma)c_3 + \frac{1}{4}(1-\gamma)^2c_1 \right],\\
r^b_6(c)&= c_6 + (1-\gamma) \left[ c_5 + \frac{1}{2}(1-\gamma)c_3 + \frac{1}{4}(1-\gamma)^2c_1 \right]. 
\end{align*}
To illustrate further this case, assume $(c,E)=((2,5,5,3,6),\,5)$ and $\gamma=0.5$. Then, $$r^b(c)=\left(1, \frac{11}{2}, \frac{11}{4}, \frac{35}{8}, \frac{59}{16}, \frac{187}{16} \right),$$ and thus the allocation is $$x^b=\left(\frac{5}{29}, \frac{55}{58}, \frac{55}{116}, \frac{175}{232}, \frac{295}{464}, \frac{935}{264} \right).$$ 

Finally, for case (c), and each $i \in \{1,\ldots, 6\}$, the allocation is given by $x_i=r_i^c(c) \frac{E}{C}$. The proportional shares $r_1^c(c)$, $r_2^c(c)$, and $r_3^c(c)$ are constructed analogously to $r_1^b(c)$ and $r_2^b(c)$ in the previous case. Agent 4 is the distinctive element in this configuration. To compute its share, the residual amounts from its two upstream agents (2 and 3) are added to its own claim, retaining a fraction $\gamma$ of the resulting total. The remaining portion is transferred to agent 5. The final shares are presented below.
\begin{align*}
r^c_1(c)&= \gamma c_1, \\
r^c_2(c)&= \gamma \left[c_2 + \frac{1}{2}(1-\gamma)c_1 \right], \\
r^c_3(c)&= \gamma \left[c_3 + \frac{1}{2}(1-\gamma)c_1 \right], \\
r^c_4(c)&= \gamma \left[c_4+ (1-\gamma)\left( c_2 + \frac{1}{2}(1-\gamma)c_1 \right) + (1-\gamma)\left( c_3 + \frac{1}{2}(1-\gamma)c_1 \right) \right]\\
&= \gamma \left[c_4+ (1-\gamma) c_2 + (1-\gamma) c_3 + (1-\gamma)^2 c_1 \right],\\
r^c_5(c)&= c_5 + (1-\gamma) \left[c_4+ (1-\gamma) c_2 + (1-\gamma) c_3 + (1-\gamma)^2 c_1 \right].
\end{align*}
To illustrate further this case, assume $(\hat{c},E) = ((2,5,5,3,6,8),\,5)$ and $\gamma=0.5$. Then, $$r^c(\hat{c})=\left(1, \frac{11}{4}, \frac{11}{4}, \frac{17}{4}, \frac{41}{4}\right),$$ and thus the allocation is $$x^c=\left(\frac{5}{21}, \frac{55}{84}, \frac{55}{84}, \frac{85}{84}, \frac{205}{84} \right).$$ 

To conclude, we mention that the previous analysis could be extended further to allow for a more general application of the geometric rules with agent-specific (rather than uniform) transfers downstream. In doing so, we would be able to accommodate other aspects that are absent from our benchmark model. For instance, suppose (as in our case study) that agents in our model represent cities. Then, one might want to consider population explicitly (if claims do not account for this aspect properly) in the allocation process. This could be done by considering a parameter $\gamma_i$ for each agent (city) $i\in N$, instead of a unique parameter $\gamma$ throughout the set. Similarly, these agent-specific parameters could also accommodate the length of the river, which is another aspect that is absent from our benchmark model.  

\section{Discussion}\label{discussion}
We have examined in this paper the design of water pollution allocation rules, building upon the river pollution claims model introduced by \cite{yang2025pollute}. In such a model, a planner sets allowable water pollution discharge amounts for each entity while considering both fairness (the standard approach) and environmental damages, where the location along the river segment determines the level of damages from that entity's discharges. In our paper, we provide (and characterize) a family of \textit{geometric} allocation rules, which further consider the level of water pollution damages based on the upstream/downstream relationship of the entities. In short, these rules allow the planner to compromise between fairness and environmental concerns by applying concatenated transfers of residual claims to downstream entities.
The family strikes a balance between the two canonical (albeit extreme) allocation rules in this setting: the \textit{fairness-blind} full transfer rule (which allocates the whole budget to the most downstream agent) and the \textit{environmentally-blind} proportional rule (which allocates the whole budget proportionally to agents' claims). 
This balance is quantified by the parameter ($\gamma$) describing the family, which represents the \textit{pollution quota} an agent retains, whereas the remaining share is transferred to the next downstream agent. 
Our axioms do not drive towards a specific value of this parameter (except for additivity and equal treatment of equal claims, which select the extreme values, $\gamma=0$ and $\gamma=1$, respectively). In other words, the combination of our axioms leaves enough room to select a variety of compromises, and further axiom(s) would be required to select a specific one. We have explored how, for instance, claims-boundedness can help in this task. 
\newpage

\cite{yang2025pollute} characterized another family, averaging between the full transfer rule and the proportional rule. 
The two families formalize (in two alternative ways) desirable compromises between environmental concerns and fairness. In our case study, averaging rules benefit the most downstream agent more than geometric rules. This seems to indicate that averaging rules prioritize environmental concerns over fairness, whereas geometric rules do the opposite. Nevertheless, to weigh these two alternative families properly, one should resort to the axiomatic analysis that supports each of them. 
As Theorem \ref{thm_geom} states, the family of geometric rules is characterized by scale invariance, upstream invariance, equal treatment of equal single polluters, top consistency and budget additivity if and only if it is a geometric rule. The family of averaging 
rules is characterized by continuity, independence of upstream null claims, budget additivity, redistribution additivity, and merging/splitting proofness (as Theorem 1 in \citet{yang2025pollute} states). Although only one axiom is shared by both characterization results, most of the remaining axioms are satisfied by all members of both families. 
This is the case of continuity, scale invariance, independence of upstream null claims, equal treatment of equal single polluters, upstream invariance and redistribution additivity.\footnote{
One might say that continuity and scale invariance play similar roles in each result, referring to operational principles. 
Likewise, independence of upstream null claims is somewhat reminiscent of equal treatment of equal single polluters and upstream invariance. 
} Thus, selecting one family over another ultimately boils down to selecting between top consistency and merging/splitting proofness (formally defined below), which are both variable-population axioms. If one prefers the former axiom, then one should endorse the family of geometric rules. If, on the other hand, one prefers the latter axiom, then one should endorse the family of averaging rules (see Table \ref{table2}). 

\begin{table}[H]
\begin{tabular}{lcc}
\toprule
Axiom & Averaging rules ($\varphi^\lambda$) & Geometric rules ($\varphi^\gamma$) \\
\midrule
Scale invariance & Y & Y$^*$ \\
Budget additivity & Y$^*$ & Y$^*$ \\
Equal treatment of equal single polluters & Y & Y$^*$ \\
Upstream invariance & Y & Y$^*$ \\
Top consistency & N & Y$^*$ \\
Continuity & Y$^*$ & Y \\
Merging/splitting proofness & Y$^*$ & N \\
Independence of upstream null claim & Y$^*$ & Y \\
Redistribution additivity & Y$^*$ & Y \\
\bottomrule
\end{tabular}
\caption{This table summarizes the axioms satisfied or violated by each family of rules. ``Y" means that the family satisfies the axiom, while ``N" that it does not. In addition, Y$^*$ means that this axiom, together with the
others marked with Y$^*$ in the same column, characterizes the corresponding rules.\label{table2}}
\end{table}

For the sake of completeness, let us formally define merging/splitting proofness here. 

\textbf{Merging/splitting proofness}.\footnote{The reader is referred to \cite{ju2007non} for a careful analysis of axioms of this sort in various settings.} For each $n>1$, let $u_n\in\mathcal{R}^{N}$ be such that $E=c_1=1$. Then, for each $i<n$,
$$
\varphi_i\left(u_n\right)+\varphi_{i+1}\left(u_n\right)=\varphi_i\left(u_{n-1}\right).
$$
The axiom applies to elementary problems in which only the most upstream agent has a positive and unitary claim. If, for any of those problems, two adjacent agents merge to be considered a single agent, the axiom prevents them from benefiting after such a move.

It is not difficult to show that the full transfer and proportional rules are the only geometric rules satisfying this axiom. Likewise, they are the only averaging rules satisfying top consistency. 

We add to the previous discussion that the family of geometric rules can also be characterized resorting only to fixed-population axioms. That is, we could obtain an alternative result to Theorem \ref{thm_geom} in which top consistency is replaced by another fixed-population axiom formalizing a related idea.\footnote{Roughly speaking, the axiom would state that if a subset of agents revisits the allocation of the total amount awarded to the subset, while assuming that the remaining agents have null claims, nothing changes. \cite{Dietzenbacher24} have recently explored a similar axiom for claims problems and \cite{Martinez2026} have done so for the adjudication of water rights in international rivers.}

\section*{Appendix A. Proofs of the results}

\subsection*{Proof of Lemma \ref{lemma}}

Let $\varphi$ and $\varphi'$ be two rules that satisfy \emph{budget additivity} and such that, for each $(c,E)\in \mathcal{R}$, $\varphi(c,E)=\varphi'(c,E)$. Let $c \in \mathbb{R}^n_+$ be given and, for each $i\in N$, define the function $f_i^c:[0,C] \longrightarrow \mathbb{R}^n_+$ as $f_i^c(E) = \varphi_i(c,E) \ge 0$. By \emph{budget additivity}, it follows that, for each $i\in N$, and for each pair $E,E' \in \mathbb{R}_+$ such that $E+E' \in [0,C]$, 
$$
f_i^c(E+E') = f_i^c(E) + f_i^c(E').
$$
This is the well-known Cauchy's equation, whose solution is linear \citep[e.g.,][]{aczel1966lectures}.  
That is, there exists $\gamma_i$, a function of the claims vector $c$, such that $\varphi_i(c,E)=f_i^c(E)= \gamma_i(c) E$. 

Now, let $(c,E) \in \mathcal{D}^{N}$. Then, for each $i\in N$,
$$
\varphi_i(c,E) = \gamma_i(c) E = \frac{E}{C} \varphi_i (c, C), 
$$
and 
$$
\varphi_i'(c,E) = \gamma_i'(c) E =  \frac{E}{C} \varphi_i' (c, C). 
$$
As $\varphi$ and $\varphi'$ coincide in $\mathcal{R}$, it follows that $\varphi_i(c, C)=\varphi_i'(c, C)$, and therefore, $\varphi_i(c, E)=\varphi_i'(c, E)$. \qed

\subsection*{Proof of Theorem \ref{thm_supergeom}}

It is straightforward to show that generalized geometric rules satisfy all the axioms in the statement. We prove the converse implication in two steps.  

\begin{itemize}
\setlength{\itemindent}{0.4cm}
\item[Step I.] Redistribution problems.

Let $N=\{1,\dots n\}$, and let $\Gamma: \mathbb{R}_+ \longrightarrow  \mathbb{R}_+$ be such that, for each $t\in\Bbb{R}_+$,
$$
\Gamma(t) =  \varphi_1((t,0,\ldots,0),t).
$$
Notice that, by definition of rule, $0 \leq \Gamma(t) \leq t$. Besides, by \emph{equal treatment of equal single polluters}, for each $i \in \{1,\ldots,n-1\}$,
$$
\varphi_i((\overbrace{0, \ldots, 0}^{i-1}, t, \overbrace{0, \ldots, 0}^{n-i}),t) = \varphi_1((t,0_{-1}),t) = \Gamma(t).
$$
Now, we proceed by induction on the number of agents in $N$. 

\begin{itemize}
\item[(i)] Case $n=1$. Let $(c,E) \in \mathcal{R}^{\{1\}}$, where $c=c_1=E$. It is obvious that $\varphi(c,E)= \varphi^\Gamma(c,E)$. 

\item[(ii)] Case $n=2$. 
Let $(c,E) \in \mathcal{R}^{\{1,2\}}$, where $c=(c_1,c_2)$ and $E=c_1+c_2$. \emph{Upstream invariance} requires that $\varphi_1(c,E) = \varphi_1((c_1,0),c_1) = \Gamma(c_1) = \varphi^\Gamma ((c_1,0),c_1)$. Therefore,
$$
\varphi(c,E)=(\Gamma(c_1), c_2 + (c_1-\Gamma(c_1))) = \varphi^\Gamma (c,E).
$$

\item[(iii)] Case $n>2$.
Assume now that the proof holds for problems in $\mathcal{R}^{N'}$ with $|N'|<n$. We show that it is also true for problems in $\mathcal{R}^N$. In fact, let $(c,E)=((c_1,\ldots,c_n),C) \in \mathcal{R}^N$. By a similar argument to the one in Case (ii), \emph{upstream invariance} implies that 
$$
\varphi_1(c,E) = \varphi_1((c_1,0,\ldots,0),c_1) = \Gamma (c_1) = \varphi^\Gamma_1(c,E).
$$ 
Notice that, as $\Gamma(c_1) \le c_1$, $\left(\left(c_2+c_1-\varphi_1(c,E),c_3,\dots, c_n \right),C-\varphi_1(c,E)\right)\in \mathcal{R}^{\{2,\ldots,n\}}$. Now, by
\emph{top consistency}, for each $k \in N \backslash \{1\}$,
\begin{align*}
\varphi_k(c,E) &=  
\varphi_k \left(\left(c_2+c_1-\varphi_1(c,E),c_3,\dots, c_n \right), C- \varphi_1(c,E) \right)\\
&= \varphi_k \left(\left(c_2+c_1-\varphi^\Gamma_1(c,E),c_3,\dots, c_n \right), C- \varphi^\Gamma_1(c,E) \right).
\end{align*}
By the induction hypothesis, 
\begin{multline*}
\varphi_k \left(\left(c_2+c_1-\varphi^\Gamma_1(c,E),c_3,\dots, c_n \right), C- \varphi^\Gamma_1(c,E) \right) = \\
\varphi^\Gamma_k \left(\left(c_2+c_1-\varphi^\Gamma_1(c,E),c_3,\dots, c_n \right), C- \varphi^\Gamma_1(c,E) \right).
\end{multline*}
And, as geometric rules satisfy \emph{top consistency}, it follows that 
$$ 
\varphi^\Gamma_k \left(\left(c_2+c_1-\varphi^\Gamma_1(c,E),c_3,\dots, c_n \right), C- \varphi^\Gamma_1(c,E) \right)=\varphi^\Gamma_k(c,E).
$$
Altogether, we have that $\varphi_k(c,E)=\varphi^\Gamma_k(c,E)$, as desired.
\end{itemize}

\item[Step II.] Full domain.

We have shown that if $\varphi$ satisfies the first four axioms, then it coincides with a geometric rule when restricted to the domain $\mathcal{R}$. We now assume that $\varphi$ also satisfies \emph{budget additivity} (last axiom). As geometric rules also satisfy this axiom, it follows from Lemma \ref{lemma} that the coincidence extends to the full domain $\mathcal{D}$. \qed
\end{itemize}

\subsection*{Proof of Theorem \ref{thm_geom}}

As geometric rules are generalized geometric rules, they all satisfy equal treatment of equal single polluters, upstream invariance, top consistency and budget additivity. It is straightforward to see that they also satisfy scale invariance. 

Conversely, let $\varphi$ be a rule that satisfies all properties in the statement. By Theorem \ref{thm_supergeom}, there exists a function $\Gamma: \mathbb{R}_+ \longrightarrow \mathbb{R}_+$ such that $\Gamma(t) \leq t$ and $\varphi \equiv \varphi^\Gamma$, where
$$
\Gamma(t) =  \varphi_1((t,0,\ldots,0),t).
$$
\emph{Scale invariance} implies that $\Gamma(t) =  \varphi_1((t,0,\ldots,0),t) = t \varphi_1((1,0,\ldots,0),1) = t \Gamma(1)$. Now, if we define the parameter $\gamma$ as $\gamma \equiv \Gamma(1)$, we obtain that $\varphi \equiv \varphi^\Gamma \equiv \varphi^\gamma$ and the proof is complete. \qed

\subsection*{Proof of Theorem \ref{thm_PROP}}

It is straightforward to show that the proportional rule satisfies all the axioms in the statement. We prove the converse implication in two steps.  

\begin{itemize}
\setlength{\itemindent}{0.4cm}
\item[Step I.] Redistribution problems.

Let $(c,E) \in \mathcal{R}^N$. We show that $\varphi_i(c,E)=c_i$ for each $i \in N$. To do so, we distinguish several cases.  
\begin{itemize}
    \item Case $i=1$. \emph{Equal treatment of equals} implies that $\varphi_1((c_1,\dots,c_1),nc_1)=c_1$. Besides, \emph{upstream invariance} requires that $\varphi_1(c,E)=\varphi_1((c_1,0,\ldots,0),c_1) = \varphi_1((c_1,\dots,c_1),nc_1)=c_1$.
    \item Case $i=2$. For each $j \in N$, let $p^{(j)} \in \mathcal{R}^N$ be the problem defined as follows:
    $$
    p^{(j)} = \left( (c_j, c_{j+1}, \ldots, c_n), E - \sum_{k<j} c_k \right).
    $$
    Note that $p^{(1)} \equiv (c,E)$. By \emph{top consistency},
    \begin{align*}
    \varphi_2(p^{(1)}) &= \varphi_1 \left( (c_2+(c_1-\varphi_1(p^{(1)})),c_3,\ldots,c_n), E-\varphi_1(p^{(1)})  \right) \\
    &= \varphi_1 \left( (c_2,c_3,\ldots,c_n), E-c_1 \right) \\
    &= \varphi_1 \left( p^{(2)} \right).
    \end{align*}
    From Case $i=1$, we already know that $\varphi_1 \left( p^{(2)} \right)=c_2$. Thus, $\varphi_2(c,E)=c_2$.
    \item Case $i=3$. By \emph{top consistency} (twice),
    \begin{align*}
    \varphi_3(p^{(1)}) &= \varphi_2(p^{(2)}) \\
    &= \varphi_2 \left( (c_2,c_3,\ldots,c_n), E-c_1 \right) \\
    &= \varphi_1 \left( (c_3+(c_2-\varphi_1(p^{(2)})),c_4,\ldots,c_n), E - c_1 - \varphi_1(p^{(2)}) \right). 
    \end{align*}
    From Case $i=2$, we know that $\varphi_1(p^{(2)}) = c_2$, and then
    $$
    \varphi_3(p^{(1)}) = \varphi_1 \left( (c_3,c_4,\ldots,c_n), E - c_1 - c_2 \right) =  \varphi_1 \left( p^{(3)} \right).
    $$
    From Case $i=1$, $\varphi_1 \left( p^{(3)} \right)=c_3$, and therefore $\varphi_3(c,E)=\varphi_3(p^{(1)})=c_3$.
\end{itemize}
By iteratively applying this process, we ultimately conclude that $\varphi(c,E) \equiv c \equiv \varphi^P(c,E)$.

\item[Step II.] Full domain.

We have shown that $\varphi$ coincides with $\varphi^P$ in the domain $\mathcal{R}$. As both $\varphi$ and $\varphi^P$ satisfy \emph{budget additivity}, by applying Lemma \ref{lemma} we conclude that such a coincidence extends to the full domain $\mathcal{D}$. \qed
\end{itemize}

\subsection*{Proof of Theorem \ref{thm_NT}}

It is straightforward to show that the full-transfer rule satisfies all the axioms in the statement. 

Conversely, let $\varphi$ be a rule satisfying all these axioms. As \emph{additivity} implies budget additivity, Theorem \ref{thm_supergeom} guarantees that $\varphi$ is a generalized geometric rule $\varphi^\Gamma$. Let us consider the problems $(c,E)=\left((t,0,\ldots,0), \frac{t}{2} \right)$ and $(\hat{c},E')=((0,1,0,\ldots,0), 1)$. On the one hand, according to the definition of generalized geometric rule, $\varphi^\Gamma_1(c,E)=\frac{\Gamma(t)}{2}$ and $\varphi^\Gamma_1(\hat{c},E')=0$. On the other hand, $\varphi^\Gamma_1(c+\hat{c},E+E')=\varphi^\Gamma_1 \left( (t,1,0,\ldots,0), \frac{t+2}{2} \right)=\frac{\Gamma(t)(t+2)}{2(t+1)}$. \emph{Additivity} requires that $\frac{\Gamma(t)}{2}=\frac{\Gamma(t)(t+2)}{2(t+1)}$, which implies that $\Gamma(t)=0$. Therefore, $\varphi \equiv \varphi^0 \equiv \varphi^{FT}$. \qed

\section{Appendix B. Excerpts from the Water Pollution Prevention and Control Law of the People's Republic of China (Amended in 2017)}

\textit{Article 2. This Law applies to the prevention and control of pollution of rivers, lakes, canals, irrigation channels, reservoirs and other surface waters and ground waters within the territory of the People's Republic of China.}

\textit{Article 5. A province, city, county or township shall establish a river chief system, and organize and lead such work as the water resource protection of rivers and lakes, administration of waters and bank lines, prevention and control of water pollution, and governance of water environment within its administrative region by degree and section.}

\textit{Article 7. The state encourages and supports the scientific and technological research on the prevention and control of water pollution, the application and promotion of advanced technologies as well as the publicity and education of water environment protection.}

\textit{Article 16 (excerpt). The planning for the prevention and control of water pollution of a river or lake across more than one county in a province, autonomous region or municipality directly under the Central Government shall be prepared by the administrative department of environmental protection under the people's government of the province, autonomous region or municipality directly under the Central Government together with the competent department of water administration at the same level in accordance with the planning for the prevention and control of water pollution of important rivers and lakes determined by the state and in light of the local situation, be submitted to the people's government of the province, autonomous region or municipality directly under the Central Government for approval and be filed with the State Council for archival purpose.}

\textit{Article 22. Enterprises, public institutions and individual industrial and commercial households which discharge water pollutants to waters shall set up pollutant discharge outlets in accordance with the laws, administrative regulations and the provisions of the administrative department for environmental protection of the State Council; if such outlets lead to rivers or lakes, they shall also abide by the provisions of the water administrative department of the State Council.}

\textit{Article 38. It is prohibited to stockpile or store solid wastes and other pollutants at bench land and bank slopes below the highest water level of rivers, lakes, canals, channels and reservoirs.}

\textit{Article 59. Vessels shall discharge oil-polluted water or domestic sewage in accordance with the standards for the discharge of pollutants by vessels. Maritime navigation vessels must abide by the standards of inland rivers for the discharge of pollutants by vessels as long as they enter inland rivers or ports.}

\newpage
\bibliography{JR_references.bib} 

@Article{Brink2012,
  author    = {Ren{\'{e}} van den Brink and Gerard van der Laan and Nigel Moes},
  journal   = {Journal of Environmental Economics and Management},
  title     = {Fair agreements for sharing international rivers with multiple springs and externalities},
  year      = {2012},
  pages     = {388--403},
  volume    = {63},
  doi       = {10.1016/j.jeem.2011.11.003},
  publisher = {Elsevier {BV}},
}

@Article{Gudmundsson2019,
  author    = {Jens Gudmundsson and Jens Leth Hougaard and Chiu Yu Ko},
  journal   = {Journal of Environmental Economics and Management},
  title     = {Decentralized mechanisms for river sharing},
  year      = {2019},
  pages     = {67--81},
  volume    = {94},
  doi       = {10.1016/j.jeem.2019.01.004},
  publisher = {Elsevier {BV}},
}

@Article{Oeztuerk2020,
  author    = {Z. Emel \"Ozt\"urk},
  journal   = {Journal of Environmental Economics and Management},
  title     = {Fair social orderings for the sharing of international rivers: A leximin based approach},
  year      = {2020},
  pages     = {102302},
  volume    = {101},
  doi       = {10.1016/j.jeem.2020.102302},
  publisher = {Elsevier {BV}},
}

@Article{EstevezFernandez21,
  author    = {Arantza Est{\'{e}}vez-Fern{\'{a}}ndez and Jos{\'{e}}-Manuel Gim{\'{e}}nez-G{\'{o}}mez and Mar{\'{\i}}a Jos{\'{e}} Sol{\'{\i}}s-Baltodano},
  journal   = {European Journal of Operational Research},
  title     = {Sequential bankruptcy problems},
  year      = {2021},
  pages     = {388--395},
  volume    = {292},
  doi       = {10.1016/j.ejor.2020.10.038},
  publisher = {Elsevier {BV}},
}

@Article{AlcaldeUnzu2020,
  author    = {Jorge Alcalde-Unzu and Mar{\'{\i}}a G{\'{o}}mez-R{\'{u}}a and Elena Molis},
  journal   = {International Journal of Game Theory},
  title     = {Allocating the costs of cleaning a river: expected responsibility versus median responsibility},
  year      = {2020},
  pages     = {185--214},
  volume    = {50},
  doi       = {10.1007/s00182-020-00746-w},
  publisher = {Springer Science and Business Media {LLC}},
}

@Article{Ansink12,
  author    = {Erik Ansink and Hans-Peter Weikard},
  journal   = {Social Choice and Welfare},
  title     = {Sequential sharing rules for river sharing problems},
  year      = {2012},
  pages     = {187--210},
  volume    = {38},
  doi       = {10.1007/s00355-010-0525-y},
  publisher = {Springer Science and Business Media {LLC}},
}

@Article{Ansink15,
  author    = {Erik Ansink and Hans-Peter Weikard},
  journal   = {Social Choice and Welfare},
  title     = {Composition properties in the river claims problem},
  year      = {2015},
  pages     = {807--831},
  volume    = {44},
  doi       = {10.1007/s00355-014-0862-3},
  publisher = {Springer Science and Business Media {LLC}},
}

@Article{beal2013,
  author    = {Sylvain Beal and Amandine Ghintran and Eric Remila and Philippe Solal},
  journal   = {International Game Theory Review},
  title     = {The river sharing problem: a survey},
  year      = {2013},
  pages     = {1340016},
  volume    = {15},
  doi       = {10.1142/s0219198913400161},
  publisher = {World Scientific Pub Co Pte Lt},
}

@Article{Hougaard17,
  author    = {Jens Leth Hougaard and Juan D. Moreno-Ternero and Mich Tvede and Lars Peter {\O}sterdal},
  journal   = {Games and Economic Behavior},
  title     = {Sharing the proceeds from a hierarchical venture},
  year      = {2017},
  pages     = {98--110},
  volume    = {102},
  doi       = {10.1016/j.geb.2016.10.016},
  publisher = {Elsevier {BV}},
}

@Article{ONeill1982,
  author  = {O'Neill, B.},
  journal = {Mathematical Social Sciences},
  title   = {{A} problem of rights arbitration from the {T}almud},
  year    = {1982},
  pages   = {345--371},
  volume  = {2},
  doi     = {10.1016/0165-4896(82)90029-4},
}

@Book{Thomson2019,
  author    = {Thomson, William},
  publisher = {Cambridge University Press},
  title     = {How to divide when there isn't enough: {F}rom {A}ristotle, the {T}almud, and {M}aimonides to the axiomatics of resource allocation},
  year      = {2019},
  isbn      = {1316646440},
  pagetotal = {504},
}

@article{thomson2012axiomatics,
  title={On the axiomatics of resource allocation: interpreting the consistency principle},
  author={Thomson, William},
  journal={Economics \& Philosophy},
  volume={28},
  number={3},
  pages={385--421},
  year={2012},
  publisher={Cambridge University Press}
}

@Article{AlcaldeUnzu2015,
  author    = {Jorge Alcalde-Unzu and Mar{\'{\i}}a G{\'{o}}mez-R{\'{u}}a and Elena Molis},
  journal   = {Games and Economic Behavior},
  title     = {Sharing the costs of cleaning a river: the upstream responsibility rule},
  year      = {2015},
  issn      = {0899-8256},
  month     = mar,
  pages     = {134--150},
  volume    = {90},
  doi       = {10.1016/j.geb.2015.02.008},
  publisher = {Elsevier BV},
}

@Article{Shapley1953,
  author  = {Shapley, L. S.},
  journal = {Annals of Mathematics Studies},
  title   = {{A} {V}alue for ${N}$-person {G}ames},
  year    = {1953},
  pages   = {307-317},
  volume  = {28},
  doi     = {10.1017/cbo9780511528446.003},
}

@Article{moreno2006impartiality,
  author    = {Moreno-Ternero, Juan D. and Roemer, John E.},
  journal   = {Econometrica},
  title     = {Impartiality, priority, and solidarity in the theory of justice},
  year      = {2006},
  number    = {5},
  pages     = {1419--1427},
  volume    = {74},
  publisher = {Wiley Online Library},
}

@Article{Dong2012,
  author    = {Dong, Baomin and Ni, Debing and Wang, Yuntong},
  journal   = {Environmental and Resource Economics},
  title     = {Sharing a polluted river network},
  year      = {2012},
  issn      = {1573-1502},
  month     = jun,
  number    = {3},
  pages     = {367--387},
  volume    = {53},
  doi       = {10.1007/s10640-012-9566-2},
  publisher = {Springer Science and Business Media LLC},
}

@Article{Gudmundsson2024,
  author    = {Gudmundsson, Jens and Hougaard, Jens Leth and Ko, Chiu Yu},
  journal   = {Management Science},
  title     = {Sharing sequentially triggered losses: {A}utomated conflict resolution through smart contracts},
  year      = {2024},
  issn      = {1526-5501},
  month     = mar,
  number    = {3},
  pages     = {1773--1786},
  volume    = {70},
  doi       = {10.1287/mnsc.2023.4772},
  publisher = {Institute for Operations Research and the Management Sciences (INFORMS)},
}

@article{ju2007non,
  title={Non-manipulable division rules in claim problems and generalizations},
  author={Ju, Biung-Ghi and Miyagawa, Eiichi and Sakai, Toyotaka},
  journal={Journal of Economic Theory},
  volume={132},
  number={1},
  pages={1--26},
  year={2007},
  publisher={Elsevier}
}

@Article{Wang2022,
  author    = {Wang, Yuntong},
  journal   = {Mathematical Social Sciences},
  title     = {The river sharing problem with incomplete information},
  year      = {2022},
  issn      = {0165-4896},
  month     = may,
  pages     = {91--100},
  volume    = {117},
  doi       = {10.1016/j.mathsocsci.2022.03.003},
  publisher = {Elsevier BV},
}

@Article{Ni2007,
  author    = {Ni, Debing and Wang, Yuntong},
  journal   = {Games and Economic Behavior},
  title     = {Sharing a polluted river},
  year      = {2007},
  issn      = {0899-8256},
  month     = jul,
  number    = {1},
  pages     = {176--186},
  volume    = {60},
  doi       = {10.1016/j.geb.2006.10.001},
  publisher = {Elsevier BV},
}

@article{yang2025pollute,
  title={How to pollute a river if you must},
  author={Yang, Yuzhi and Ansink, Erik and Gudmundsson, Jens},
  journal={Journal of Environmental Economics and Management},
  volume    = {130},
  pages={103105},
  year={2025},
  publisher={Elsevier}
}

@Article{Dietzenbacher24,
  author    = {Dietzenbacher, Bas and Tamura, Yuki and Thomson, William},
  journal   = {Social Choice and Welfare},
  title     = {Partial-implementation invariance and claims problems},
  year      = {2024},
  issn      = {1432-217X},
  volume    = {63},
  number    = {1},
  pages     = {203--229},
  doi       = {10.1007/s00355-024-01528-z},
  publisher = {Springer Science and Business Media LLC},
}

@book{aczel1966lectures,
  title={Lectures on functional equations and their applications},
  author={Acz{\'e}l, J{\'a}nos},
  year={1966},
  publisher={Academic press}
}

@article{moulin2000priority,
  title={Priority rules and other asymmetric rationing methods},
  author={Moulin, Herv{\'e}},
  journal={Econometrica},
  volume={68},
  number={3},
  pages={643--684},
  year={2000},
  publisher={Wiley Online Library}
}

@article{chun1988proportional,
  title={The proportional solution for rights problems},
  author={Chun, Youngsub},
  journal={Mathematical Social Sciences},
  volume={15},
  number={3},
  pages={231--246},
  year={1988},
  publisher={Elsevier}
}

@Article{hougaard2022optimal,
  author    = {Hougaard, Jens Leth and Moreno-Ternero, Juan D. and {\O}sterdal, Lars Peter},
  journal   = {Management Science},
  title     = {Optimal management of evolving hierarchies},
  year      = {2022},
  number    = {8},
  pages     = {6024--6038},
  volume    = {68},
  publisher = {INFORMS},
}

@Article{Ambec02,
  author    = {Stefan Ambec and Yves Sprumont},
  journal   = {Journal of Economic Theory},
  title     = {Sharing a river},
  year      = {2002},
  pages     = {453--462},
  volume    = {107},
  doi       = {10.1006/jeth.2001.2949},
  publisher = {Elsevier {BV}},
}

@Article{Xie2023,
  author    = {Xie, Rui and Zhang, Jiahuan and Tang, Chuan},
  journal   = {Resource and Energy Economics},
  title     = {Political connection and water pollution: New evidence from Chinese listed firms},
  year      = {2023},
  issn      = {0928-7655},
  month     = aug,
  pages     = {101390},
  volume    = {74},
  doi       = {10.1016/j.reseneeco.2023.101390},
  publisher = {Elsevier BV},
}

@Article{Chakraborti2022,
  author    = {Chakraborti, Lopamudra and Shimshack, Jay P.},
  journal   = {Resource and Energy Economics},
  title     = {Environmental disparities in urban Mexico: Evidence from toxic water pollution},
  year      = {2022},
  issn      = {0928-7655},
  month     = feb,
  pages     = {101281},
  volume    = {67},
  doi       = {10.1016/j.reseneeco.2021.101281},
  publisher = {Elsevier BV},
}

@Article{Arguedas2020,
  author    = {Arguedas, Carmen and Cabo, Francisco and Martín-Herrán, Guiomar},
  journal   = {Journal of Environmental Economics and Management},
  title     = {Enforcing regulatory standards in stock pollution problems},
  year      = {2020},
  issn      = {0095-0696},
  month     = mar,
  pages     = {102297},
  volume    = {100},
  doi       = {10.1016/j.jeem.2019.102297},
  publisher = {Elsevier BV},
}

@Book{Thomson2023,
  author    = {Thomson, William},
  publisher = {Springer International Publishing},
  title     = {The Axiomatics of Economic Design, Vol. 1: An Introduction to Theory and Methods},
  year      = {2023},
  isbn      = {9783031293986},
  doi       = {10.1007/978-3-031-29398-6},
  issn      = {2197-8530},
  journal   = {Studies in Choice and Welfare},
}

@Article{Gudmundsson2025,
  author    = {Gudmundsson, Jens and Hougaard, Jens L. and Moreno-Ternero, Juan D. and {\O}sterdal, Lars Peter},
  journal   = {Games and Economic Behavior},
  title     = {Optimizing successive incentives: {R}ewarding the past or motivating the future?},
  year      = {2025},
  issn      = {0899-8256},
  month     = oct,
  pages     = {10-29},
  volume    = {153},
  doi       = {10.1016/j.geb.2025.05.006},
  publisher = {Elsevier BV},
}

@Article{Hougaard2012,
  author    = {Hougaard, Jens Leth and Moreno-Ternero, Juan D. and {\O}sterdal, Lars Peter},
  journal   = {Journal of Mathematical Economics},
  title     = {A unifying framework for the problem of adjudicating conflicting claims},
  year      = {2012},
  issn      = {0304-4068},
  month     = mar,
  number    = {2},
  pages     = {107--114},
  volume    = {48},
  doi       = {10.1016/j.jmateco.2012.01.004},
  publisher = {Elsevier BV},
}

@Article{Martinez2025,
  author    = {Mart\'inez, Ricardo and Moreno-Ternero, Juan D.},
  journal   = {Environmental and Resource Economics},
  title     = {The allocation of riparian water rights},
  year      = {2025},
  issn      = {1573-1502},
  month     = nov,
  number    = {12},
  pages     = {3841--3872},
  volume    = {88},
  doi       = {10.1007/s10640-025-01044-3},
  publisher = {Springer Science and Business Media LLC},
}

@Article{Liu2005,
  author    = {Liu, Jianguo and Diamond, Jared},
  journal   = {Nature},
  title     = {China’s environment in a globalizing world},
  year      = {2005},
  issn      = {1476-4687},
  month     = jun,
  number    = {7046},
  pages     = {1179--1186},
  volume    = {435},
  doi       = {10.1038/4351179a},
  publisher = {Springer Science and Business Media LLC},
}

@Article{Rice2017,
  author    = {Rice, Jacelyn and Westerhoff, Paul},
  journal   = {Nature Geoscience},
  title     = {High levels of endocrine pollutants in US streams during low flow due to insufficient wastewater dilution},
  year      = {2017},
  issn      = {1752-0908},
  month     = jul,
  number    = {8},
  pages     = {587--591},
  volume    = {10},
  doi       = {10.1038/ngeo2984},
  publisher = {Springer Science and Business Media LLC},
}

@Article{Hougaard2013,
  author    = {Hougaard, Jens Leth and Moreno-Ternero, Juan D. and {\O}sterdal, Lars Peter},
  journal   = {Social Choice and Welfare},
  title     = {Rationing in the presence of baselines},
  year      = {2013},
  issn      = {1432-217X},
  month     = apr,
  number    = {4},
  pages     = {1047--1066},
  volume    = {40},
  doi       = {10.1007/s00355-012-0664-4},
  publisher = {Springer Science and Business Media LLC},
}

@Article{Martinez2026,
  author    = {Mart\'inez, Ricardo and Moreno-Ternero, Juan D.},
  journal   = {arXiv preprint arXiv:2601.04150},
  title     = {The geometric adjudication of water rights in international rivers},
  year      = {2026},
  copyright = {Creative Commons Attribution 4.0 International},
  publisher = {arXiv},
}
\bibliographystyle{mystyle3}

\end{document}